\documentstyle[preprint,eqsecnum,aps,epsf]{revtex}

\begin{document}


\newtheorem{prop}{Proposition}

\preprint{YITP-96-36}
\title{
Can the entanglement entropy be the origin of \\
black-hole entropy ?
}
\author{
Shinji Mukohyama${}^\dagger$,  
Masafumi Seriu${}^{\dagger, \ddagger}$\\ 
and \\
Hideo Kodama${}^\dagger$
}
\address{
Yukawa Institute for Theoretical Physics, 
Kyoto University${}^\dagger$ \\
Kyoto 606-01, Japan \\
and\\
Institute of Cosmology, Physics \& Astronomy Department,
Tufts University${}^\ddagger$ \\
Medford, Massachusetts 02155, USA
}
\date{December 25, 1996}

\maketitle


\begin{abstract} 
Entanglement entropy is often speculated as a strong candidate for the
origin of the black-hole entropy. To judge whether this speculation is 
true or not, it is effective to investigate the whole structure of 
thermodynamics obtained from the entanglement entropy, rather than
just to examine the apparent structure of the entropy alone or to
compare it with that of the black hole entropy. It is because entropy
acquires a physical significance only when it is related to the energy
and the temperature of a system. From this point of view, we construct
a `thermodynamics of entanglement' by introducing an entanglement
energy and compare it with the black-hole thermodynamics. We consider
two possible definitions of entanglement energy. Then we construct 
two different kinds of thermodynamics by combining each of these
different definitions of entanglement energy with the entanglement
entropy. We find that both of these two kinds of thermodynamics show
significant differences from  the black-hole thermodynamics if no
gravitational effects are taken into account. These differences are in
particular highlighted in the context of the third law of
thermodynamics. Finally we see how inclusion of gravity alter the
thermodynamics of the entanglement. We give a suggestive argument that
the thermodynamics of the entanglement behaves like the black-hole
thermodynamics if the gravitational effects are included
properly. Thus the entanglement entropy passes a non-trivial check to
be the origin of the black-hole entropy. 
\end{abstract}

\vfill


\section{Introduction}
\label{sec:Introduction}
Understanding the origin of the black-hole entropy is one of the most
fascinating problems in black-hole physics~\cite{WALD,BH-entropy}. The
concept of the black-hole entropy traces back to the work by
Bekenstein, who pointed out that the behavior of the basic physical
quantities describing stationary black holes has an analogous
structure to that of ordinary thermodynamical
systems~\cite{Bekenstein}. This thermodynamical structure inherent in
the black-hole theory is usually called the `black-hole
thermodynamics'. In analogy with ordinary material systems, then, it
is natural to expect that the black-hole entropy  comes from
microscopic degrees of freedom of a system including a black
hole. This suggests that quantum theory of gravity should inevitably
take part in the black-hole thermodynamics. In this sense,
understanding the origin of the black-hole entropy shall provide us
with important information on quantum gravity. This is one among  the
several reasons why the black-hole entropy needs to be understood at
the fundamental level. 

Let us recall basic properties of the  black-hole thermodynamics by
taking a simple example. We consider  the   one-parameter family of
Schwarzschild black holes parameterized by the mass $M_{BH}$. Here and
throughout this paper, we assume that the relation analogous to the
first law of thermodynamics holds for a black-hole system. In the
present example, there is only one parameter $M_{BH}$ characterizing a
black hole. Therefore, this relation should be of the simplest form
%
\begin{equation}
 dE_{BH} = T_{BH}dS_{BH}\ \ \ ,	
\label{eqn:1st-law}
\end{equation}
where $E_{BH}$, $S_{BH}$ and $T_{BH}$ are quantities that are
identified with the energy, the entropy and the temperature of a black
hole, respectively. The relation Eq.(\ref{eqn:1st-law}) is called the
first law of the black-hole
thermodynamics~\cite{Bekenstein,Wald1993}. Thus, if two of the 
quantities $E_{BH}$, $S_{BH}$ and $T_{BH}$ are given,
$Eq.(\ref{eqn:1st-law})$ determines the remaining quantity. For
simplicity, let us call this procedure of defining energy, entropy
and temperature which satisfy the first law (Eq.(\ref{eqn:1st-law}))
the `construction of thermodynamics'. 

In the present example, $M_{BH}$ is the only parameter characterizing
the family of black holes. Therefore the simplest combination which
yields the dimension of energy is 
%
\begin{equation}
 E_{BH} \equiv M_{BH}c^2\ \ \ .
\label{eqn:E-BH}
\end{equation}
This is the energy of the black hole.

There is also a natural choice for
$T_{BH}$~\cite{Hawking1975}. Hawking showed 
that a black hole with surface gravity $\kappa$ emits thermal
radiation of a matter field (which plays the role of a thermometer) at 
temperature $k_BT_{BH}=\hbar\kappa /2\pi c$. Moreover one can show
that if any matter field in a thermal-equilibrium state at any
temperature is scattered by a black hole, it goes to another
thermal-equilibrium state at a temperature closer to
$T_{BH}$~\cite{detailed-balance}. Thus it is natural to define the
temperature of a Schwarzschild black hole with mass $M_{BH}$ by
%
\begin{equation}
 k_BT_{BH} = \frac{\hbar c^3}{8\pi GM_{BH}}\ \ \ ,	
\label{eqn:T-BH}
\end{equation}
since $\kappa =c^4/4GM_{BH}$~\cite{WALD}. 

From Eqs.(\ref{eqn:1st-law})-(\ref{eqn:T-BH}),
we can construct the thermodynamics for the Schwarzschild black
holes. Thus we get an expression for  $S_{BH}$ given by
%
\begin{equation}
 S_{BH} = \frac{k_B c^3}{4\hbar G}A+C\ \ \ ,	
\label{eqn:S-BH}
\end{equation}
where $A \equiv 16\pi G^2 M_{BH}^2 / c^4$ is the area of 
the event horizon and $C$ is some constant. Since a value of $C$ is
not essential in our discussions, we shall set hereafter
%
\begin{equation}
 C = 0\ \ \ .
\label{eqn:constants}
\end{equation}
It is well-known that classically the area of the event horizon does
not decrease in time  just as the ordinary thermodynamical
entropy. The result Eq.(\ref{eqn:S-BH}) looks reasonable in this
sense. Indeed this observation was the original motivation for the
introduction of the black-hole thermodynamics~\cite{Bekenstein}.
However, it is not clear to what extent $S_{BH}$ is related with the
information as the ordinary thermodynamical entropy is. At this stage
we would like to point out that the third law of thermodynamics does
not hold for a black hole irrespective of the choice of the value for
$C$ as is seen from Eqs.(\ref{eqn:T-BH}) and (\ref{eqn:S-BH}). We
shall come back to this point in \S\ref{subsection:discrepancy}. In
any case, understanding the origin 
of $S_{BH}$ is an important problem  in black-hole physics. 

There have been many attempts to  understand the origin of the
black-hole entropy~\cite{BH-entropy}. Among them we shall concentrate
only on the 
so-called  entanglement entropy~\cite{BLLS,Srednicki,CW,FN,Sent}. The
aim of this paper is to judge whether entanglement entropy can be
regarded as the origin of black-hole entropy. For this purpose it is
effective to investigate the whole structure of the thermodynamics
obtained from the entanglement entropy, rather than just to examine
the apparent structure of the entropy alone. Thus we shall construct
the `thermodynamics of entanglement' and compare it with the
black-hole thermodynamics. As is expected by the above example of the
black-hole thermodynamics, we have to define either energy or
temperature to construct the thermodynamics of entanglement. Combining
it with the entanglement entropy, which is already at
hand~\cite{BLLS,Srednicki,CW,FN}, the other is automatically defined by
means of Eq.(\ref{eqn:1st-law}). In this paper we shall choose the
option to  define the entanglement energy firstly, deriving the
entanglement temperature afterwards. We shall consider two possible
definitions of entanglement energy. Therefore we can construct two
different kinds of thermodynamics by combining each of these
definitions of entanglement energy with the entanglement
entropy. We show that neither of these thermodynamics is compatible
with the black-hole thermodynamics if no gravitational effects are
taken into account. After that, we see how inclusion of gravity alter
the thermodynamics of the entanglement. We give a suggestive argument
that they have a common behavior if gravitational effects are taken
into account properly. Thus the entanglement entropy passes a
non-trivial check to be the origin of the black-hole entropy. 

The rest of the paper is organized as follows. In section
\ref{sec:Sent} we review the concept of the entanglement entropy. In
section \ref{sec:Eent} we propose two natural definitions of
entanglement energy and present general formulas for calculating the
energy. In section \ref{sec:example} explicit expression for the
entanglement energy are derived for some tractable models with the
help of the formulas prepared in section
\ref{sec:Eent}. In section \ref{sec:discussion} we construct
the thermodynamics of entanglement and compare it with the black hole
thermodynamics from various angles. Section \ref{sec:summary} is
devoted to the summary of our  results.


\section{Entanglement entropy}	
\label{sec:Sent}
In this section we review the definition and basic properties of the
entanglement entropy. 

\subsection{Definition of the entanglement entropy}
\label{subsection:Senta}
Here we give a general definition of entanglement entropy, since it is
not usually made clear in the literature. 

Let $\cal U$ be a Hilbert space constructed from two Hilbert spaces
$\cal V$ and $\cal W$ as 
%
\begin{equation}
{\cal U} = {\cal V} \bar{\otimes} {\cal W} \ \ \ ,
\label{eqn:Hilbert}
\end{equation}
where $\bar{\otimes}$ denotes a tensor product followed by a suitable
completion. We call an element $u \in {\cal U}$  {\it prime} if $u$
can be written as $u = v \otimes w$ with $v \in {\cal V}$ and 
$w\in {\cal W}$. For example, 
$u= v_1 \otimes w_1 + 2 v_1 \otimes w_2 
+ v_2 \otimes w_1 + 2 v_2 \otimes w_2$ 
is prime since $u$ can be represented as 
$u = (v_1 + v_2) \otimes (w_1 + 2w_2) $.  
On the other hand 
$u= v_1 \otimes w_1 + v_2 \otimes w_2 $ 
is not prime if neither $v_1$ and $v_2$ nor $w_1$ and $w_2$ are
linearly dependent. The entanglement entropy 
$S_{ent} : {\cal U} \rightarrow {\bf R}_+ 
=\{ {\rm non-negative\  real\ numbers} \}$ 
defined below can be regarded as a measure of the non-prime nature of
an element of ${\cal U} = {\cal V} \bar{\otimes} {\cal W}$. 
 
First of all, from an element $u$ of ${\cal U}$ with unit norm we
construct an operator $\rho$ 
(`density operator') by
%
\begin{equation}
 \rho v = (u,v)u \quad {}^\forall v\in{\cal U},
\label{eqn:rho}
\end{equation}
where $(u,v)$ is the inner product which is antilinear with respect to 
$u$. In this context $\rho$ represents a `pure state'.
 
From $\rho$ we define another operator (`reduced density operator') 
$\rho_{\cal W}$ by
%
\begin{equation}
 \rho_{\cal W}y = \sum_{i,j}f_j(e_i\otimes f_j,\rho e_i\otimes y)
	\quad {}^\forall y\in{\cal W},
\label{eqn:reduced}
\end{equation}
where $\{ e_i\}$ and $\{ f_j\}$ are orthonormal bases of ${\cal V}$
and ${\cal W}$ respectively. Note that 
%
\begin{equation}
 {\rm Tr}_{\cal W}\left(\rho_{\cal W}A\right) =
	{\rm Tr}_{\cal U}\left(\rho\  1\otimes A\right)
\end{equation}
for an arbitrary bounded operator $A$ on ${\cal W}$.
  
Finally we define the entanglement entropy with respect to $\rho$ as 
%
\begin{equation}
  S_{ent} \left[ \rho \right] 
     \equiv   -k_B{\rm Tr}_{\cal W} \left[ \rho_{\cal W} 
                      \ln \rho_{\cal W} \right]\ \ \ .
\label{eqn:entropy}
\end{equation}
We can totally exchange the roles played by $\cal V$ and $\cal W$ in
Eq.(\ref{eqn:reduced}) and Eq.(\ref{eqn:entropy}). The entanglement
entropy is so defined as to be invariant under the exchange of $\cal
V$ and $\cal W$ when $\rho$ corresponds to a pure state, i.e., when
$\rho$ is given by Eq.(\ref{eqn:rho}). (See {\it Appendix} for
the proof of this property.) 

As a simple example, let us consider spin states for a system
consisting of an electron and a proton. We take 
${\cal V}= \{ | u \rangle , | d \rangle \}$ for an electron and 
${\cal W}= \{ | U \rangle , | D \rangle \}$ for a proton, where `$u$'
and `$U$' are for `up', while `$d$' and `$D$'  are  for `down'. Then
${\cal U} = {\cal V} \otimes {\cal W}$ is spanned by 
%
\[
 \{ | u \rangle \otimes | U \rangle , 
     | u \rangle \otimes | D \rangle , 
     | d \rangle \otimes | U \rangle , 
     | d \rangle \otimes | D \rangle \} \ \ \ .  
\]
Now let us consider a state 
%
\begin{eqnarray*}
   | \phi \rangle &=& (\alpha | u \rangle + \beta | d \rangle)
                    \otimes (A| U \rangle + B| D \rangle)\ \ \ , \\
     &{}& |\alpha|^2 +|\beta|^2 = |A|^2 + |B|^2 =1 \ \ \ ,
\end{eqnarray*}
which is clearly a prime state. 
According to Eq.(\ref{eqn:reduced}), we then get
%
\[
 \rho_e = 
       \left(
           \begin{array}{cc}
              |\alpha|^2      &  \alpha \beta^* \\
	      \alpha^* \beta  &  |\beta|^2     
           \end{array}	
      \right) \ \ \ .
\]
Here `$e$' is for `electron'.
By a suitable diagonalization of this matrix, it is easy to see that
$S_{ent} =0$. We can exchange the roles between `electron' and
`proton': we then get
%
\[
 \rho_p = 
       \left(
           \begin{array}{cc}
              |A|^2      &  A B^* \\
	      A^* B  &  |B|^2     
           \end{array}	
      \right) \ \ \ ,
\]
(`$p$' is for `proton') which again leads to $S_{ent} =0$. 

On the contrary, an $s$-state 
%
\begin{eqnarray*}
   | \phi' \rangle &=& \gamma | u \rangle \otimes | D \rangle 
                + \delta | d \rangle \otimes | U \rangle \ \ \ , \\
     &{}& |\gamma|^2 +|\delta|^2 =1, \gamma \delta \neq 0 
\end{eqnarray*}
is not a prime state. For this state the reduced density operators are 
given by
%
\[
 \rho_e = 
       \left(
           \begin{array}{cc}
              |\gamma|^2      &  0  \\
	         0  &  |\delta|^2     
           \end{array}	
      \right) \ \ \ , 
 \rho_p = 
       \left(
           \begin{array}{cc}
              |\delta|^2      &  0  \\
	         0  &  |\gamma|^2     
           \end{array}	
      \right) \ \ \ .
 \]
Therefore we get 
$S_{ent} = -k_B (|\gamma|^2 \ln |\gamma|^2 
                   + |\delta|^2 \ln |\delta|^2) > 0$.

\subsection{Relevance to black hole entropy}	
In the case of black-hole physics, the presence of
the event horizon causes a natural decomposition of a Hilbert space
$\cal F$ of all states of matter fields to a tensor product of the
state spaces inside and outside a black hole as
%
\begin{equation}
 {\cal F} = {\cal F}_1 \bar{\otimes} {\cal F}_2 \ \ \ .		
 \label{eqn:F=F1*F2}
\end{equation}
For example, let us  take  a scalar field. 
We can suppose that its one-particle Hilbert space $\cal H$ is
decomposed as
%
\begin{equation}
 {\cal H} = {\cal H}_1 \oplus {\cal H}_2 \ \ \ ,
	\label{eqn:H=H1+H2}
\end{equation}
where ${\cal H}_1$ is a space of mode functions with
supports inside the horizon and ${\cal H}_2$ is a space of mode
functions with supports outside the horizon. Then we can
construct new Hilbert spaces (`Fock spaces') ${\cal F}$, ${\cal F}_1$
and ${\cal F}_2$ from ${\cal H}$, ${\cal H}_1$ and ${\cal H}_2$,
respectively, as 
%
\begin{eqnarray}
 {\cal F} & \equiv &
    \mbox{\boldmath C} \oplus {\cal H} \oplus 
    \left({\cal H}\bar{\otimes} {\cal H}\right)_{sym}
    \oplus \cdots \ \ \ ,	
    \nonumber	\\
 {\cal F}_1 & \equiv &
    \mbox{\boldmath C} \oplus {\cal H}_1 \oplus 
    \left({\cal H}_1\bar{\otimes} {\cal H}_1\right)_{sym}
    \oplus \cdots \ \ \ ,	
    \nonumber	\\
 {\cal F}_2 & \equiv &  
    \mbox{\boldmath C} \oplus {\cal H}_2 \oplus 
    \left({\cal H}_2 \bar{\otimes} {\cal H}_2 \right)_{sym} 
    \oplus \cdots \ \ \ ,	
    \label{eqn:Fock}
\end{eqnarray}
where $(\cdots)_{sym}$ denotes the symmetrization. Now these three
Hilbert spaces satisfy the relation (\ref{eqn:F=F1*F2}). Hence the
entanglement entropy $S_{ent}$ is defined by the procedure given at
the beginning of this section (Eqs.(\ref{eqn:rho})-(\ref{eqn:entropy})) 
for each state in ${\cal F}$.
 
The entanglement entropy $S_{ent}$ originates from a tensor product
structure of the Hilbert space  as Eq.(\ref{eqn:F=F1*F2}), which is
caused by the existence of the  boundary between two regions (the
event horizon) through Eq.(\ref{eqn:H=H1+H2}). Furthermore the
symmetric property of $S_{ent}$ between $\cal V$ and $\cal W$
mentioned before also suggests that $S_{ent}$ is
related with a boundary between two regions. In fact $S_{ent}$ turns
out to be proportional to the area of such a boundary (a model for the
event horizon) in simple models (see the next subsection). In view of 
Eq.(\ref{eqn:S-BH}) with Eq.(\ref{eqn:constants}), thus, the
entanglement entropy has a nature similar to the black hole entropy. 

The relevance of the entanglement entropy to the black hole entropy is 
also suggested by the following observation. 
Let us consider a free scalar field on a background geometry
describing a gravitational collapse to a black hole. We compare the
black hole entropy and the entanglement entropy for this system. We
begin with the black hole entropy. In the initial region of the
spacetime, there is no horizon and the entropy around this region can
be regarded as zero. In 
the final region, on the other hand, there is an event  horizon so
that the black-hole possesses non-zero entropy. As for the
entanglement entropy, the existence of the event horizon naturally
divides the Hilbert space ${\cal F}$ of all states of the scalar field
into ${\cal F}_1\bar{\otimes}{\cal F}_2$. Thus according to the
argument in the previous subsection, the scalar field in some pure
state possesses non-zero entanglement entropy. In this manner, we
observe that the black-hole entropy and the entanglement entropy come
from the same origin, i.e. the existence of the event horizon. This is
the reason why the entanglement entropy is regarded as one of the
potential candidates for the origin of the black-hole entropy.

\subsection{Simple models}	
\label{subsection:BHcase}
The relation between the entanglement entropy and the black hole
entropy was analyzed in terms of simple tractable models by
\cite{BLLS} and \cite{Srednicki}. They considered a free scalar field
on a flat spacelike hypersurface $\Sigma={\rm R}^3$ embedded in a
4-dimensional Minkowski spacetime, and calculated the entanglement
entropy for a division of $\Sigma$ into two regions $\Sigma_1$ and
$\Sigma_2$ with a common boundary $B$. 
Here $\Sigma_1$,  $\Sigma_2$ and $B$ are, respectively,  the
models of the interior, the exterior of the black holes and the
horizon. Ref.\cite{BLLS} chooses $B$ to be a 2-dimensional flat
surface, and the matter state to be the ground state, showing that the
resulting entanglement entropy becomes proportional to the area of
$B$. Ref.\cite{Srednicki} chooses $B$ to be a $2$-sphere in ${\rm
R}^3$, and chooses $\Sigma_1$ and $\Sigma_2$ to be the interior and
the exterior of the sphere. The matter state is chosen to be the
ground state. Then it is shown that the resulting entanglement entropy
is again proportional to the area of $B$. 

Both of the results can be expressed as 
%
\begin{equation}
 S_{ent}[\rho_0] = \frac{k_B  {\cal N}_S}{a^2}A \ \ \ , 
\label{eqn:Sent-result}
\end{equation}
where $\rho_0$ is the  ground-state density matrix, $A$ is the area of
the boundary, $a$ is a cutoff length, and ${\cal N}_S$ is a
dimensionless numerical constant of order unity. This coincides with 
(\ref{eqn:S-BH}) and (\ref{eqn:constants}) if the cut-off length $a$
is chosen as 
%
\begin{equation}
 a = \sqrt{  { {4{\cal N}_S \hbar G} \over  {c^3} } }
  = 2  \sqrt{{\cal N}_S}\ l_p  \ \ \ ,	\label{eqn:a=lpl}
\end{equation}
where $l_p$ is the Planck length. Here note that $a$ depends
only on the Planck length. 

In this paper we adopt the same simple models to construct
thermodynamics of entanglement, and to discuss its relevance to the
black hole thermodynamics. There the relation (\ref{eqn:a=lpl}) and
the subsequent comment will play an important role.

\section{Entanglement energy}	
\label{sec:Eent}
In this section we define the  entanglement energy to construct
the thermodynamics of entanglement. We give two possible
definitions of entanglement energy. The difference between them 
comes from the difference in the way to formulate the
reduction of a system (caused by, for instance, the formation of an
event horizon). In the first definition (\S\ref{subsection:Eenta} and
\S\ref{subsection:Eentb}), we assume that the state of a total system
undergoes a change in the course of the reduction of the system (so
that the density matrix of the system changes actually), while
operators are regarded as unchanged. In the second definition
(\S\ref{subsection:Eentc} and \S\ref{subsection:Eentd}), on the
contrary, we assume that some operators drop out from the set of all
observables, while the state is regarded  as unchanged. Since at
present we cannot  judge whether and which one of these treatments
reflects the true process of reduction, the best way is to
investigate both options. As we shall see in \S\ref{sec:discussion},
the universal behavior of the thermodynamics of entanglement does not
depend on the choice of the entanglement energy. 

To make our analysis a concrete one, we apply these definitions to the
tractable models given in \S\ref{subsection:BHcase}. Let us consider a
system described by a Fock space $\cal F$ constructed from a
one-particle Hilbert space $\cal H$ in the previous section. Let
$H_{tot}$ be a total  Hamiltonian acting  on $\cal F$. We assume that
the Hamiltonian $H_{tot}$ is naturally decomposed as 
%
\begin{equation}
 H_{tot} = H_1 + H_2 + H_{int} \ \ \ ,		
\label{eqn:Htot=H1+H2+Hint}
\end{equation}
where $H_1$ and $H_2$ are parts acting on ${\cal F}_1$ and 
${\cal F}_2$, respectively, and $H_{int}$ is a part representing the
interaction of two regions.

\subsection{The first definition of the entanglement energy}
\label{subsection:Eenta}
First let us consider the case in which the total density operator
$\rho$ actually changes to the product of reduced density operators of 
each subsystems, $\rho_1$ and $\rho_2$, (when, for instance,  an event
horizon is formed), while the observables remain unchanged. 

$\rho$ reduces to $\rho^I$ given  by 
%
\begin{equation}
  \rho^I = \rho_1 \otimes \rho_2\ \ \ .	
\label{eqn:rhoI}
\end{equation}
It is easy to see that the entropy associated with this density matrix
becomes 
%
\begin{equation}
 -k_B{\rm Tr} \left[\rho^I\ln\rho^I\right] =
	 S_{ent}[\rho ] + S'_{ent}[\rho ],	
	 \label{eqn:Sent-tot}
\end{equation}
where $S_{ent}[\rho ]$ and $S'_{ent}[\rho ]$ are entanglement entropy
obtained through $\rho_1$ and $\rho_2$, respectively. $S_{ent}[\rho ]$ 
and $S'_{ent}[\rho ]$ are identical if $\rho$ is a pure state (see
the argument below Eq. (\ref{eqn:entropy})).

It is clear that  the partial systems labeled by `1' and `2' can be
treated symmetrically: the symmetric property of the entanglement
entropy for a pure state shows that it measures the entanglement
between ${\cal F}_1$ and ${\cal F}_2$, so that it is symmetrical in
nature. Accordingly, one can exchange the suffices `1' and `2' in the
above formulas.

Since we are assuming that the observables do not change, we are led
to the following first definition of entanglement energy:
%
\begin{equation}
 E_{ent}^I \equiv {\rm Tr}\left[:H_{tot}:\rho^I\right] \ \ \ , 
  \label{eqn:Eent1}
\end{equation}
where $:\ -\ :$ denotes the usual normal ordering (a subtraction of
the ground state energy).

\subsection{Formula of $E_{ent}^I$ for the ground state}
\label{subsection:Eentb}
What we should do next is to calculate $E_{ent}^I$ explicitly by
choosing $\rho$ as the ground state of $H_{tot}$. We consider a free
scalar field and discretize it with some spatial separation for
regularization. Since the system thus obtained is 
equivalent to a set of harmonic oscillators, in this section, we give
a formula of $E_{ent}^I$ for the ground state of coupled harmonic
oscillators. In the next section we calculate $E_{ent}^I$ explicitly
by using the formula.

Let us consider a system of coupled harmonic oscillators  
$\left\{ q^A\right\}$ $(A=1,\cdots,N)$ described by the Lagrangian, 
%
\begin{equation}
 L = \frac{a}{2}\delta_{AB}\dot{q}^A\dot{q}^B - 
     \frac{1}{2}V_{AB}q^A q^B \ \ \ .
\label{eqn:Lagrange}
\end{equation}
Here $\delta_{AB}$ is Kronecker's delta symbol\footnote{
From now on, we choose the units $\hbar =c=1$ and apply
Einstein's summation convention unless otherwise stated.}; 
$V$ is a real-symmetric, positive-definite matrix which does not
depend on $\left\{ q^A\right\}$. We have introduced $a$$(>0)$ as a
fundamental length characterizing the system.\footnote{
Thus $\left\{ q^A\right\}$ are treated as dimension-free quantities in
the present units.}
The corresponding Hamiltonian becomes 
%
\begin{equation}
 H_{tot} = \frac{1}{2a}\delta^{AB}p_A p_B + 
           \frac{1}{2}V_{AB}q^A q^B \ \ \ ,		
\label{eqn:Htot}
\end{equation}
where $p_A=a\delta_{AB}\dot{q}^B$ is the canonical momentum conjugate
to $q^A$. 

Firstly we calculate the wave function 
$\langle\left\{q^A\right\} |0\rangle$ of the ground state
$|0\rangle$. Note that Eq.(\ref{eqn:Htot}) can be written as
%
\begin{equation}
 H_{tot} = \frac{1}{2a}\delta^{AB} 
           \left(p_A+iW_{AC}q^C\right)\left(p_B-iW_{BD}q^D\right) +
           \frac{1}{2a}{\rm Tr} W
\end{equation}
by using the commutation relation 
$\left[ q^A,p_B\right] =i\delta^A_B$. Here $W$ is a symmetric matrix
satisfying $(W^2)_{AB} = a V_{AB}$. The ambiguity in sign is fixed by
requiring $W$ to be positive definite. Now 
$\langle\left\{q^A\right\} | 0\rangle$ is given as the solution to
%
\begin{equation}
 \left(\frac{\partial}{\partial q^A}+W_{AB}q^B\right)
	\langle\left\{q^A\right\} | 0\rangle = 0 \ \ \ ,
\end{equation}
since $p_A$ is expressed as $-i\frac{\partial}{\partial q^A}$.  The
solution is 
%
\begin{equation}
 \langle\left\{q^A\right\} | 0\rangle =
      \left(\det\frac{W}{\pi}\right)^{1/4} 
      \exp\left( -\frac{1}{2}W_{AB}q^A q^B \right)\ \ \ ,
\end{equation}
which is normalized with respect to the standard Lebesgue measure 
$dq^1 \cdots dq^N$. The corresponding density matrix $\rho_0$
corresponding to this ground state is represented as 
%
\begin{eqnarray}
 \langle\left\{q^A\right\}| \rho_0 |\left\{q'^B\right\}\rangle
& = &
      \langle\left\{q^A\right\}|0\rangle	
      \langle 0|\left\{q'^B\right\}\rangle \nonumber	\\
& = & 
      \left( \det \left(\frac{W}{\pi}\right) \right)^{1/2} 
      \exp\left[ -\frac{1}{2}W_{AB}
       \left(q^A q^B + {q'}^A q'^B\right)\right].
 \label{eqn:ex-rho0}
\end{eqnarray}

Now we split $\left\{ q^A\right\}$ into two
subsystems, $\left\{ q^a\right\}$ $(a=1,\cdots,n)$ and 
$\left\{ q^{\alpha}\right\}$ $(\alpha =n+1,\cdots,N)$. (We assign the
labels `1' and `2' to the former and the latter subsystems,
respectively.) Then we obtain the reduced density matrix associated
with the subsystem 2 (the subsystem 1),  by taking the  partial trace
of $\rho_0$ w.r.t. the subsystem 1 (the subsystem 2): 
%
\begin{eqnarray}
 \langle\left\{q^{\alpha}\right\}| \rho_2 
                |\left\{q'^{\beta}\right\}\rangle
& = & 
      \int \prod_{c=1}^{n}dq^{c}
      \langle\left\{q^a,q^{\alpha}\right\}| \rho_0
            | \left\{q^b,q'^{\beta}\right\}\rangle  \nonumber \\
& = & 
      \left(\det\frac{D'}{\pi}\right)^{1/2}
      \exp\left[ -\frac{1}{2}D'_{\alpha\beta}
          \left( q^{\alpha}q^{\beta}+q'^{\alpha}q'^{\beta}\right)
               \right]	\nonumber  \\
& & 
\times\exp\left[ -\frac{1}{4}\left(B^{T}A^{-1}B\right)_{\alpha\beta}
		(q-q')^{\alpha}(q-q')^{\beta}
               \right]
\end{eqnarray}
and
%
\begin{eqnarray}
 \langle\left\{q^a\right\}| \rho_1 |\left\{q'^b\right\}\rangle
& = & 
      \int\prod_{\gamma=n+1}^{N}dq^{\gamma}
      \langle\left\{q^a,q^{\alpha}\right\}| \rho_0
             | \left\{q'^b,q^{\beta}\right\}\rangle  \nonumber \\
& = & 
      \left(\det\frac{A'}{\pi}\right)^{1/2}
      \exp\left[ -\frac{1}{2}A'_{ab}
          \left( q^a q^b + q'^a q'^b \right)
               \right]	\nonumber	\\
& & 
\times\exp\left[ -\frac{1}{4}\left(B D^{-1}{B^T}\right)_{ab}
	          (q-q')^a(q-q')^b
               \right],
\end{eqnarray}
where $A$, $B$, $D$, $A'$ and $D'$ are defined by
%
\begin{eqnarray}
 \left( W_{AB}\right) & = & \left(
 \begin{array}{cc}
	A_{ab}			& B_{a \beta} \\
	(B^{T})_{\alpha b}	& D_{\alpha \beta}	
 \end{array}	\right)\ \ \ ,		\nonumber \\
 A'     & = & A - BD^{-1}B^{T}\ \ \ ,	\nonumber \\
 D'     & = & D - B^{T}A^{-1}B\ \ \ .	
 \label{eqn:ABD}  
\end{eqnarray}
(The superscript $T$ denotes transposition.) Here note that $A^T=A$
and $D^T=D$. Thus $\rho^I$ defined by Eq.(\ref{eqn:rhoI}) is
represented as 
%
\begin{eqnarray}
 \langle\left\{q^A\right\}| \rho^I |\left\{q'^B\right\}\rangle
& = &
      \left(\det\frac{M}{\pi}\right)^{1/2}
      \exp\left[ -\frac{1}{2}M_{AB}
          \left( q^A q^B + q'^A q'^B \right)
               \right]	\nonumber	\\
& & 
\times\exp\left[ -\frac{1}{4}N_{AB}(q-q')^A(q-q')^B\right]
\ \ \ ,  
\end{eqnarray}
where 
%
\begin{eqnarray}
 \left( M_{AB}\right) & = & \left(
 \begin{array}{cc}
	A'_{ab}	& 0			\\
	0	& D'_{\alpha\beta}
 \end{array}	\right)\ \ \ ,	\nonumber\\
 \left( N_{AB}\right) & = & \left(
 \begin{array}{cc}
	\left( BD^{-1}B^{T}\right)_{ab} & 0	\\
	0 & \left(B^{T}A^{-1}B\right)_{\alpha\beta}
 \end{array}	\right)\ \ \ .
\end{eqnarray}

We can diagonalize $M$ and $N$ simultaneously by the following
non-orthogonal transformation: 
%
\begin{equation}
 q^A \to \tilde{q}^A \equiv \left( UM^{1/2}\right)^A_{\ B} q^B\ \ \ ,
\label{eqn:change}
\end{equation}
where $U$ is a real orthogonal matrix  satisfying
%
\begin{eqnarray}
 M^{-1/2} N M^{-1/2} 
 &=& 
 U^{T}\lambda U	\ \ \ , 	\nonumber\\
 \lambda
& = & 
 \left( \begin{array}{cccccc}
	\lambda_1 & & \\
	& \lambda_2 & \\
	& & \ddots
               \end{array}\right).
\end{eqnarray}
Now in terms of  $\left\{\tilde{q}^A\right\}$,  
   $H_{tot}$ is  represented as 
%
\begin{equation}
 H_{tot} =  
 -\frac{1}{2a}\left( UMU^{T}\right)^{AB}
	\left( \frac{\partial}{\partial \tilde{q}^A} - 
		\tilde{W}_{AC}\tilde{q}^C\right)
	\left( \frac{\partial}{\partial \tilde{q}^B} + 
		\tilde{W}_{BD}\tilde{q}^D\right)
        + \frac{1}{2a}{\rm Tr} W \ \ \ , 
\end{equation}
thus, 
\begin{equation}
 :H_{tot}:= 
 -\frac{1}{2a}\left( UMU^{T}\right)^{AB}
	\left( \frac{\partial}{\partial \tilde{q}^A} - 
		\tilde{W}_{AC}\tilde{q}^C\right)
	\left( \frac{\partial}{\partial \tilde{q}^B} + 
		\tilde{W}_{BD}\tilde{q}^D\right)\ \ \ ,
\end{equation}
where 
%
\begin{equation}
 \tilde{W} \equiv UM^{-1/2}WM^{-1/2}U^{T} \ \ \ .
\end{equation}
Hence the density matrix $\rho^I$ is expressed in terms of 
$|\left\{ \tilde{q}^A \right\} \rangle $ as\footnote{
Einstein's summation convention is not applied to
Eq.(\ref{eqn:ex-rhoI}). }
%
\begin{equation}
 \langle\left\{\tilde{q}^A\right\}| \rho^I 
	|\left\{\tilde{q}'^B\right\}\rangle =  
 \prod_{C=1}^N \pi^{-1/2}
   \exp\left[ -\frac{1}{2}\left\{(\tilde{q}^C)^2 + (\tilde{q}'^C)^2 
				\right\}
                 -\frac{1}{4}
                  \lambda_C (\tilde{q}^C-\tilde{q}'^C)^2
          \right] \ \ \ .	
\label{eqn:ex-rhoI}
\end{equation}
This density matrix is normalized with respect to the measure 
$d\tilde{q}^1\cdots d\tilde{q}^N$. 

Now it is easy to calculate the entanglement energy. First the matrix
components of $:H_{tot}:\rho^I$ with respect to $\{\tilde{q}^A\}$ are
given by
%
\begin{eqnarray}
 \langle\left\{\tilde{q}^A\right\} | :H_{tot}:\rho^I | 
				\left\{\tilde{q}^B\right\}\rangle
& = & 
 -\frac{1}{2a}
 \Big\{
	\left[ UMU^{T}-lUM^{-1/2}VM^{-1/2}U^{T}\right]_{AB}
		\tilde{q}^A \tilde{q}^B 	\nonumber\\
&{}&   +{\rm Tr}\left[ W- N/2 - M \right]
 \Big\}
 \prod_{C=1}^{N}\pi^{-1/2}
 \exp\left[ -(\tilde{q}^C)^2\right] \ \ \ .
\end{eqnarray}
Hence from the definition (\ref{eqn:Eent1}) the entanglement energy
$E_{ent}^I$ is expressed as 
%
\begin{eqnarray}
 E_{ent}^I 
& = &
 \int(\prod_{C=1}^{N}d\tilde{q}^C)
   \langle \left\{ \tilde{q}^C  \right\} | :H_{tot}:\rho^I | 
	\left\{ \tilde{q}^B \right\} \rangle \nonumber \\
& = & 
 \frac{1}{4a}{\rm Tr}\left[ lVM^{-1}+M+N-2W\right] \ \ \ .
\end{eqnarray}
Here we have used the formula
$\int d{\vec x} \ {\vec x}\cdot{\cal A} {\vec x}\ 
           \exp[-{\vec x}\cdot {\vec x}] 
  = {1\over 2} \pi^{N/2} {\rm Tr} {\cal A}$, 
where $N$ is the dimension of ${\vec x}$.
With the help of the identity 
$ {\rm Tr}\left[ M+N\right] = {\rm Tr} A + {\rm Tr} D 
= {\rm Tr} W $, 
we finally arrive at the following formula for $E_{ent}^I$
%
\begin{eqnarray}
 E_{ent}^I 
& = & 
 \frac{1}{4a}{\rm Tr}\left[ a V M^{-1} -W \right]   
 \label{eqn:Eent1'} \nonumber \\
& = &
  \frac{1}{4}{\rm Tr}\left[ V(M^{-1}-W^{-1})\right]	\nonumber \\
& = & 
 -\frac{1}{2}{\rm Tr}\left[ V_{int}^{T}\tilde{B}\right]\ \ \ .	
\label{eqn:ex-Eent1}
\end{eqnarray}
Here $V_{int}$ is a block in the matrix $V$ given by 
%
\begin{equation}
 \left( V_{AB}\right)  =  \left(
 \begin{array}{cc}
	V^{(1)}_{ab}	&	\left({V_{int}}\right)_{a\beta}  \\
	({V_{int}}^{T})_{\alpha b} & V^{(2)}_{\alpha\beta}
 \end{array}	\right)	\ \ \ ,
 \label{eqn:V}
 \end{equation}
and $\tilde{B}$ is a block  in the matrix  $W^{-1}$
given by 
%
\begin{equation}
 W^{-1}  =  
 \left(
 \begin{array}{cc}
	(A'^{-1})^{ab}	& \tilde{B}^{a\beta}	\\
	(\tilde{B}^{T})^{\alpha b} & (D'^{-1})^{\alpha\beta}	
 \end{array}	
 \right) \ \ \ .	
 \end{equation}
It is easy to see that 
%
\begin{equation}
\tilde{B}^{a\beta} 
= -\left( {A'}^{-1} B D^{-1} \right)^{a\beta}
= -\left( {A}^{-1} B {D'}^{-1} \right)^{a\beta} \ \ \ .
\end{equation}


\subsection{The second definition of the entanglement energy}
\label{subsection:Eentc}
The second definition of entanglement energy follows
when we regard that the operators connecting the two subsystems drop
out from the set of observables (when, for instance, an event horizon
is formed), while the state of the system is regarded as
unchanged. To be more precise, we assume that  $H_1$ and $H_2$ remain
to be observables but that $H_{int}$ is no longer an observable (see
Eq.(\ref{eqn:Htot=H1+H2+Hint})). In this case it is natural to define
the entanglement energy by 
%
\begin{equation}
 E_{ent}^{II} \equiv
	{\rm Tr}\left[ (:H_1: + :H_2:)\rho\right],
\label{eqn:Eent2}
\end{equation}
where the two normal orderings mean to subtract the minimum
eigenvalues of $H_1$ and $H_2$ respectively.

\subsection{Formula of $E_{ent}^{II}$ for the ground state}
\label{subsection:Eentd}
For the ground state of the system analyzed in
\S\ref{subsection:Eentb}, we now evaluate the entanglement energy in the sense of Eq.(\ref{eqn:Eent2}).

Firstly we divide  the Hamiltonian (\ref{eqn:Htot}) into three terms
as Eq.(\ref{eqn:Htot=H1+H2+Hint}) (see Eq.(\ref{eqn:V})): 
%
\begin{eqnarray}
 H_1 
& \equiv & 
 \frac{1}{2a}\delta^{ab}p_a p_b + \frac{1}{2}{V^{(1)}}_{ab}
                 q^a q^b	\nonumber\\
& = & 
 \frac{1}{2a}\delta^{ab} 
	\left(p_a+iw^{(1)}_{ac}q^c\right)
	\left(p_b-iw^{(1)}_{bd}q^d\right) +
           \frac{1}{2a}{\rm Tr} w^{(1)} \ \ \ ,    \nonumber\\
 H_2 
& \equiv &
 \frac{1}{2a}\delta^{\alpha\beta}p_{\alpha} p_{\beta} + 
	\frac{1}{2}{V^{(2)}}_{\alpha\beta}
	     q^{\alpha} q^{\beta}          \nonumber\\
& = & 
 \frac{1}{2a}\delta^{\alpha\beta} 
	\left(p_{\alpha}+iw^{(2)}_{\alpha\gamma}q^{\gamma}\right)
	\left(p_{\beta}-iw^{(2)}_{\beta\delta}q^{\delta}\right) +
           \frac{1}{2a}{\rm Tr} w^{(2)} \ \ \ ,     \nonumber\\
 H_{int} 
 & \equiv & H_{tot} - H_1 -H_2     \nonumber\\
 &=& V_{int \  a \beta} q^a q^\beta \ \ \ ,
\end{eqnarray}
where $w^{(1)}$ and $w^{(2)}$ are, respectively, the positive
square-roots of $aV^{(1)}$ and $aV^{(2)}$. Although there exists
freedom in the way of the division, the above division seems to be the 
most natural one. Here and throughout this paper we adopt it. 

Now it is convenient to  diagonalize $W$ in (\ref{eqn:ex-rho0}) by a
non-orthogonal transformation, 
%
\begin{equation}
 q^A \to \bar{q}^A \equiv \delta ^{AB} 
      { \left( W^{1/2} \right)_{BC}  }q^C\ \ \ .
\end{equation}
In terms of  $\left\{\bar{q}^A\right\}$, the density matrix $\rho_0$
corresponding to the ground state of $H_{tot}$ is written as 
%
\begin{equation}
\langle\left\{\bar{q}^A\right\}| \rho_0
		|\left\{\bar{q}'^B\right\}\rangle
= \prod_{C=1}^{N}\pi^{-1/2}\exp\left[ -\frac{1}{2}
	\left( (\bar{q}^C)^2 + (\bar{q}'^C)^2\right)
	\right]\ \ \ . 
\end{equation}
On the other hand the operator $:H_1:+:H_2:$ is written as 
%
\begin{equation}
 :H_1: + :H_2:\  
 =  -\frac{1}{2a} \delta ^{AC} \delta ^{BD} W_{CD}
 \left(
   \frac{\partial}{\partial\bar{q}^A}-\bar{w}_{AE}\bar{q}^E
   \right)
 \left(
 \frac{\partial}{\partial\bar{q}^B}+\bar{w}_{BF}\bar{q}^F
 \right) \ \ \ ,
\end{equation}
where $\bar{w}$ is defined by
%
\begin{eqnarray}
 \bar{w} & \equiv &  
       \delta_{AC} 
               \left( W^{-1/2} w W^{-1/2} \right)^{CD} 
                        \delta_{DB} \ \ \ ,    \nonumber\\
 \left( w_{AB}\right) & \equiv &  \left(
 \begin{array}{cc}
	{w^{(1)}}_{ab}		&	0	\\
	0		& 	{w^{(2)}}_{\alpha \beta}
 \end{array}	\right) \ \ \ .
\end{eqnarray}

From these expressions we obtain
%
\begin{eqnarray}
 \langle\left\{\bar{q}^A\right\}| (:H_1:+:H_2:)\rho_0
		|\left\{\bar{q}^B\right\}\rangle
& = & 
 \frac{1}{2a}\Bigl\{
	\left[(\bar{w}+1)W(\bar{w}-1)\right]_{AB}\bar{q}^A\bar{q}^B
		\nonumber	\\
& &  - {\rm Tr}\left[ W(\bar{w}-1)\right]\Bigr\}
 \prod_{C=1}^{N}\pi^{-1/2}\exp\left[ 
	-(\bar{q}^C)^2\right].
\end{eqnarray}

Hence we arrive at the following expression $E_{ent}^{II}$ for $\rho_0$
%
\begin{eqnarray}
 E_{ent}^{II}
& = & 
 \int (\prod_{C=1}^N d\bar{q}^C)\langle\left\{\bar{q}^A\right\} |
	(:H_1:+H_2:)\rho_0|\left\{\bar{q}^B\right\}\rangle\nonumber\\
& = & 
 \frac{1}{4a}{\rm Tr}\left[ w^2W^{-1}-W\right]
	- \frac{1}{2a}{\rm Tr}\left[ w-W\right].
\end{eqnarray}
With the help of the relation 
${\rm Tr} [ w^2W^{-1}] = {\rm Tr}[ aVM^{-1}]$ 
which follows from the definitions of $w$ and $M$, this formula is
simplified as
%
\begin{eqnarray}
 E_{ent}^{II} 
& = &
 \frac{1}{4a}{\rm Tr}\left[ aVM^{-1} -W \right]
 -\frac{1}{2a}{\rm Tr}\left[ w-W\right] \nonumber	\\
& = & 
 E_{ent}^I -\frac{1}{2a}{\rm Tr}\left[ w-W\right],
\label{eqn:ex-Eent2}
\end{eqnarray}
where Eq.(\ref{eqn:Eent1'}) has been used to obtain the last line.


\section{Explicit evaluation of the entanglement energy for some
tractable models in flat spacetime}
\label{sec:example}
With the help of the formulas derived in the previous section, we  now
calculate $E_{ent}^I$ and $E_{ent}^{II}$ explicitly to construct 
two kinds of the thermodynamics of entanglement for the simple models
discussed in \S\ref{subsection:BHcase}.

We consider a free scalar field $\phi$ on 4-dimensional Minkowski
space ${\cal M}_4$, described by the action\footnote{
Here we follow the sign-convention of $diag(-,+,+,+)$.}
%
 \begin{equation}
 S = \int_{{\cal M}_4} 
 \left\{
 -{1\over 2}(\partial \phi)^2 + 
 {1\over 2} m^2 \phi^2 \right\}
  \ \ \ .
  \label{eqn:action}
 \end{equation}
  Now we  divide the   spatial section 
  ${\rm R}^3$ of ${\cal M}_4$ 
into two disconnected 
regions $\Sigma_1$ and $\Sigma_2$ by a suitable 
2-dimensional surface $B$.  We consider two cases: 
One is the choice $B={\rm R}^2$ and the other is $B={\rm S}^2$. We
calculate $E_{ent}^I$ and $E_{ent}^{II}$ 
for each case, getting four results in total.
Stating the results first, 
they are  summarized   
in the universal form\footnote{
We have recovered $\hbar$ and $c$ in Eq.(\ref{eqn:Eent-result}).}
%
\begin{equation}
 E_{ent} = {{\cal N}_E}{{\hbar c} \over {a^3}}A,	
 \label{eqn:Eent-result}
\end{equation}
where $A$ is the area of $B$, $a(>0)$ is a cutoff length and
${\cal N}_E$ is a dimensionless numerical constant.

\subsection{Entanglement energy for the case of $B={\rm R}^2$ }
\label{subsection:half}
First we take $B={\rm R}^2$. Without loss of generality the
resulting two half-spaces are represented as 
$\Sigma_1=$ $\{(x_1,x_2,x_3):$ $x_1>0\}$ and 
$\Sigma_2=$ $\{(x_1,x_2,x_3):$ $x_1<0\}$~\cite{BLLS}.

Here some comments are in order. Since all the degrees of freedom on
and across $B$, which is infinite, contribute to the entanglement
energy, a suitable cut-off length $a(>0)$
should be introduced to avoid the ultra-violet divergence. For the
same reason, the infra-red divergence is also anticipated in
advance, since $B$ is non-compact in this model. The latter is taken
care of by considering the massive case since the inverse of the mass 
characterizes a typical size of the spreading of the field. Clearly
$a$ should be taken short enough in the unit of the Compton length of
the field, $m^{-1}$, to obtain meaningful results. Therefore we shall
only pay attention to the leading order in the limit $ma\to 0$ in the
course of calculation as well as in the final results. These
remarks are valid in any model of this type, and the same remarks
apply to the case of the entanglement entropy,
too~\cite{BLLS,Srednicki}.

In order to calculate $E_{ent}^I$ for the present case, we first note
that the term $ V_{AB}q^A q^B$ in Eq.(\ref{eqn:Lagrange}) corresponds
to the expression 
$\int\left[( {\vec{\nabla}} \phi )^2 + m^2\phi^2  \right] d^3 x$ 
read off from Eq.(\ref{eqn:action}), which defines an operator
$V(x,y)$ acting on a space 
${\cal W}=\left( \left\{ \phi (\cdot) \right\}, d^3 x \right)$. 
In order to use  the formula (\ref{eqn:ex-Eent1}), thus, we need the
positive square-root and the inverse of $aV$. For this purpose, it is
convenient to work in the momentum representation of 
$\cal W$~\cite{BLLS} given by 
%
\begin{eqnarray}
 \phi (\vec{x}) &=& \int \frac{d^3 k}{(2\pi)^3} \ 
                 \phi_{\vec{k}} 
                 \exp[i \vec{k} \cdot \vec{x}] \ \ \ , \nonumber\\
 \phi_{\vec{k}} &=& \int d^3 x \ 
                 \phi (\vec{x}) 
                 \exp[-i \vec{k} \cdot \vec{x}] \ \ \ . 
\end{eqnarray}
  The results are  
%
\begin{eqnarray}
 V(\vec{x},\vec{y})
& = & 
	\int\frac{d^3 k}{ (2\pi )^3}(\vec{k}^2+m^2)
	\exp [i\vec{k} \cdot (\vec{x}-\vec{y})]\ \ \ ,\nonumber\\
 W^{-1}(\vec{x},\vec{y}) 
& = &
	\int_{{\rm R}^3} \frac{d^3 k}{(2\pi )^3}
	 (\vec{k}^2  +  m^2)^{-1/2}
	 \exp [i\vec{k} \cdot(\vec{x}-\vec{y})]\ \ \ .
\end{eqnarray}
Note that both $V(\vec{x},\vec{y})$ and 
$W^{-1}(\vec{x},\vec{y})$ are 
symmetric under the exchange of $\vec{x}$ and $\vec{y}$. 
(The cut-off must preserve this property.) 
Now the formula (\ref{eqn:ex-Eent1}) gives 
%
\begin{eqnarray}
 E_{ent}^{I} & = & 
 -\frac{1}{2}\int_{y_1 <-a}d^3 y \int_{x_1 >a}d^3 x
	\int_{|k_1|<a^{-1}}\frac{d^3 k}{(2\pi)^3}
	    ({\vec{k}}^2  +  m^2)
	\exp[i\vec{k}\cdot(\vec{y}-\vec{x})]\nonumber\\
& & \times	\int_{|k'_1|<a^{-1}}\frac{d^3 k'}{(2\pi)^3}
	    ({\vec k}^{'2}  +  m^2)^{-1/2}
	\exp[i\vec{k}' \cdot (\vec{x}-\vec{y})] \ \ \ ,
\end{eqnarray}
where, as discussed above, 
a cut-off length $a$ was introduced in the integral. 

Since the integrand is invariant under the translation along $B$, the
integral with respect to $x_2$ and $x_3$ yields a divergent factor 
$A=\int_{R^2}dx_2dx_3$. Clearly this factor should be interpreted as
the area of $B$. If this divergent integral $A$ is factored out, we
obtain the following convergent expression for $E_{ent}^{I}$:
%
\begin{eqnarray}
 E_{ent}^{I} 
& = &  
 -\frac{A}{2} \int_{-\infty}^{-a}dy_1 
              \int_{a}^{\infty}dx_1 
	\int\frac{d^2 k_\parallel}{(2\pi)^2}
	\int_{-a^{-1}}^{a^{-1}}\frac{dk_1}{2\pi}
	\int_{-a^{-1}}^{a^{-1}}\frac{d{k'}_1}{2\pi}	\nonumber\\
& & \times
	(\vec{k_\parallel}^2 + {k_1}^2 + m^2)
	(\vec{k_\parallel}^2 + {{k'}_1}^2 + m^2)^{-1/2}
   \exp[i(k_1 - {k'}_1)(y_1 -x_1 )]\ \ \ .
\end{eqnarray}
Here $\vec{k}_\parallel$ is a 2-vector lying along $B$ 
and $k_1$, ${k'}_1$ are components normal to $B$  
(if we make an obvious identification of ${\rm R}^3$ with 
its Fourier space). 
Let us change   
the  variables from $x_1$ and $y_1$ to 
$z\equiv x_1 -y_1$ and $u \equiv (x_1 +y_1 )/2$. 
Then $z$ and $u$ take values in the range 
$z \leq 2a$ and $-({z \over 2} -a) \geq u \geq ({z \over 2}-a)$, 
respectively. Hence the integration with respect to $u$ yields 
%
\begin{eqnarray}
 E_{ent}^{I} 
& = &  
 -\frac{A}{2}\int_{2a}^{\infty}dz(z-2a)
	\int \frac{d^2 k_\parallel}{(2\pi)^2}
	\int_{-a^{-1}}^{a^{-1}}\frac{dk_1}{2\pi}
	\int_{-\infty}^{\infty}\frac{dk'_1}{2\pi}	\nonumber\\
& & \times	({\vec{k}_\parallel}^2 + {k_1}^2 + m^2)
	({\vec{k}_\parallel}^2 + {{k'}_1}^2 + m^2)^{-1/2}
	\exp[-i({k_1}-{{k'}_1})z]	             \nonumber\\
& = & 
 -\frac{A}{2}\int_{2a}^{\infty}dz(z-2a)
	\int_{-a^{-1}}^{a^{-1}}\frac{dk_1}{2\pi}   \cos({k_1} z)
	\int_m ^\infty  
	    \frac{d\kappa}{2\pi}\kappa (\kappa^2+{k_1}^2)\nonumber\\
& & \times	\int_{-\infty}^{\infty}
	   \frac{d{k'}_1}{2\pi}
	(\kappa^2 + {{k'}_1}^2)^{-1/2} \cos({k'}_1 z)
\end{eqnarray}
in the leading order, where $\kappa$ is defined by
$\kappa^2 = {\vec{k}_\parallel}^2  +  m^2$. 
Here note that in this expression, the integration with respect to 
${k'}_1$ followed by that with respect to $\kappa$ leads to an
infra-red divergence if we set $m=0$, in accordance with 
our discussion at the beginning of this subsection. 

Now let us recollect some  formulas with the 
modified Bessel functions~\cite{Bessel}:
%
\begin{eqnarray}
 K_0(x) & = &
	\int_{0}^{\infty}dt\frac{\cos t}{\sqrt{t^2+x^2}}\ \ \ , \nonumber\\
 \int_{x_0}^{\infty}dx\ xK_0(x) & = &
	x_0 K_1(x_0)\ \ \ , \nonumber\\
 \int_{x_0}^{\infty}dx \ x^3 K_0(x) & = &
	x_0^3 K_1(x_0)+2x_0^2 K_2(x_0)\ \ \ .
\end{eqnarray}
With the help of  these formulas
 $E_{ent}^I$ is written as
%
\begin{eqnarray}
 E_{ent}^I &=&
 -\frac{A}{2} \int_{2a}^{\infty}dz (z-2a)
    \int_{-a^{-1} }^{a^{-1}}\frac{dk_1}{2\pi}
	\cos({k_1} z) \nonumber	\\
& & \times \frac{1}{2\pi^2}\left[
         \frac{m}{z} ({k_1}^2 + m^2) K_1(mz)  
         +  2 \frac{m^2}{z^2} K_2(mz) \right]\ \ \ , \nonumber \\
& = & 
 \frac{A}{4\pi^3 a^3}
  [ \alpha_1(ma) + \alpha_2(ma)+ \alpha_3(ma)]\ \ \ 
 \label{eqn:resultA1}
\end{eqnarray}
in the leading order. Here we have introduced 
%
\begin{eqnarray}
 \alpha_1(x)
     & \equiv &
      -x \int_2^\infty  d\xi \frac{\xi -2}{\xi^4}   K_1(x\xi )
	\left[ 2\xi \cos\xi  + (\xi^2 - 2 ) \sin\xi \right]
                                          \ \ \ ,  \nonumber\\
 \alpha_2(x)
     & \equiv &
     -2x^2 \int_2^\infty  d\xi \frac{\xi -2}{\xi^3} K_2(x\xi )
              \sin\xi   \ \ \ ,                     \nonumber\\
 \alpha_3(x)
     & \equiv &
     -x^3 \int_2^\infty d\xi \frac{\xi -2}{\xi^2} K_1(x\xi )
              \sin\xi \ \ \ .
\end{eqnarray}
A numerical evaluation shows
%
\[
 \left[ \alpha_1(x) + \alpha_2(x) + \alpha_3(x) \right] \sim 0.05
   \ \ {\rm as} \ \      x \to 0 \ \ \ .
\]
This result is of the form of 
Eq.(\ref{eqn:Eent-result}) with   
${\cal N}_E\sim 0.05/4\pi^3 \approx 4.0 \times 10^{-4}$ 
in the limit $ma \to 0$.

In order to calculate $E_{ent}^{II}$ by the 
formula (\ref{eqn:ex-Eent2}) we use the expression for $w$,
%
\begin{equation}
 w(\vec{x} , \vec{y}) = 
	\int \frac{d^3 k}{( 2\pi )^3}( \vec{k}^2 + m^2)^{1/2}
	\left\{ \theta (x_1) \theta (y_1) +
		\theta (-x_1)\theta (-y_1) \right\}
	\exp[i\vec{k} \cdot (\vec{x} -\vec{y} )]\ \ \ .
\end{equation}
From this it follows that 
%
\begin{eqnarray}
 W(x,y)-w(x,y) & = & 
	\int \frac{d^3 k}{(2\pi )^3}(\vec{k}^2 + m^2)^{1/2}
	\left\{ \theta (x_1) \theta (-y_1) +
		\theta (-x_1) \theta (y_1) \right\}\nonumber\\
& & \times	\exp[i\vec{k} \cdot (\vec{x} -\vec{y} )]\ \ \ .
\end{eqnarray}
Taking the trace of this expression, we get  
%
\begin{eqnarray}
{\rm Tr} (W-w) & = & \int d^3 x \int d^3 y 
\delta^3(\vec{x} -\vec{y} )
	\left[ W(\vec{x}, \vec{y}) - w(\vec{x}, \vec{y}) 
	\right]	\\
 & = & A\int\frac{d^3 k}{( 2\pi )^3}(\vec{k}^2 + m^2)^{1/2} \nonumber\\
& & \times	\left[\int dx_1 dy_1 \delta ( x_1 - y_1 )
	\left\{\theta (x_1) \theta (-y_1) +
	\theta (-x_1) \theta (y_1) \right\} \right]\nonumber\\
 & = & 0\ \ \ .
\end{eqnarray}
Therefore we get\footnote{
If we adopt another regularization scheme with the same cut-off length
$a$, the result may change. However, the change is in sub-leading order.}
%
\begin{equation}
 E_{ent}^{II} = E_{ent}^{I} \approx \frac{0.05A}{4\pi^3a^3}\ \ \ .
\end{equation}


\subsection{Entanglement energy for the case of $B={\rm S}^2$}
\label{subsection:sphere}
Next we consider the case $B= {\rm S}^2$, a sphere with radius $R$.
This is the same model as in the calculation of 
the entanglement entropy in Ref.\cite{Srednicki}. 
 
It should be noted that in this model, we can put $m=0$ from the
beginning without being bothered by the infra-red divergence. 
This is because $B$ is compact as is discussed  at the beginning of the
previous subsection. Hence we simply put $m=0$ in this subsection. 

By introducing the polar coordinates, $\phi (\vec{x})$ is expanded by
the spherical harmonics as
%
\begin{equation}
 \phi (r, \theta, \psi) =\sum_{l,m}  \frac{\phi_{lm}(r)}{r} 
          Z_{lm}(\theta,\psi )\ \ \ ,
\end{equation}
where $Z_{lm}$ and $Z_{l,-m}$ are the real parts of the spherical
harmonics $Y_{lm}$ and $Y_{l,-m}$, respectively. 
In terms of $\phi_{lm}$ the potential term in Eq.(\ref{eqn:action}) is 
written as 
%
\begin{equation}
 \int (\nabla\phi )^2d^3x 
  = \sum_{l,m}\int_{0}^{\infty}
            \left[ 
              r^{2}
              \left\{ \frac{\partial}{\partial r}
               \left( \frac{\phi_{lm}(r)}{r} \right) \right\}^{2}
              + \frac{l(l+1)}{r^{2}} \phi_{lm}^{2}(r)
            \right] dr.
\label{eqn:pot}
\end{equation}

Now we introduce a cut-off scale $a$ as in 
the previous model to take care of the ultra-violet divergence. 
For that purpose we divide the radial coordinate $r$ into a lattice
as $r_A = an$ $(n=1,2,\cdots,N)$, with the identification $R \equiv
(n_B+1/2)a$ for some $n_B$. It is understood that the limit
$N\to\infty$ is taken in the final results. 
As is discussed at the beginning of the previous subsection, 
the entanglement energy is carried by the modes around $B$, as the
term `entanglement' implies. We thus need to introduce the cut-off
scale $a$ only in the $r$-direction. Under this regularization, the
right-hand side of Eq.(\ref{eqn:pot}) (corresponding to
$V_{AB}q^{A}q^{B}$ in Eq.(\ref{eqn:Lagrange})) turns to 
%
\[
\frac{1}{a}\sum_{l,m}\sum_{n=1}^N\left[
      \left( n+\frac{1}{2}\right)^{2} \left( 
      \frac{\phi_{lm,n}}{n}-\frac{\phi_{lm,n+1}}{n+1}
      \right)^{2} + 
      \frac{l(l+1)}{n^{2}}{\phi_{lm,n}}^{2}\right]\ \ \ , 
\]
where we have imposed the boundary condition 
$\phi_{lm,N+1}\equiv 0$. Thus $V$ is written as the direct sum
%
\[
 V = \mathop{\oplus}_{l,m}V^{(l,m)},
\]
where $V^{(l,m)}$ is the $N\times N$ matrix given by
%
\begin{eqnarray}
 \left( V^{(l,m)}_{AB}\right) & = & \frac{2}{a}\left( 
               \begin{array}{cccccc}
                   \Sigma^{(l)}_1 & \Delta^{(l)}_1 & & & & \\
         \Delta^{(l)}_1 & \Sigma^{(l)}_2 & \Delta^{(l)}_2 & & & \\
       & \ddots & \ddots & \ddots & & \\
       & & \Delta^{(l)}_{n-1} & \Sigma^{(l)}_n & \Delta^{(l)}_n & \\
       & & & \ddots & \ddots & \ddots 
               \end{array}
             \right) \ \ \ , \nonumber\\
   \Sigma^{(l)}_n
          & = & 1+\frac{1}{4n^2}+\frac{l(l+1)}{2n^2}\ \ \ ,\nonumber\\
   \Delta^{(l)}_n
          & = & -\frac{(n+1/2)^2}{2n(n+1)} \ \ \ .
\end{eqnarray}

Hence $E_{ent}$ is written as
%
\[
 E_{ent} = \sum_{l=0}^\infty
  (2l+1)E_{ent}^{(l)}\ \ \ ,
\]
where $E_{ent}^{(l)}$ is defined by (\ref{eqn:ex-Eent1}) or
(\ref{eqn:ex-Eent2}) with $V$ replaced by $V^{(l,m)}$.

\begin{figure}
\centerline{\epsfbox{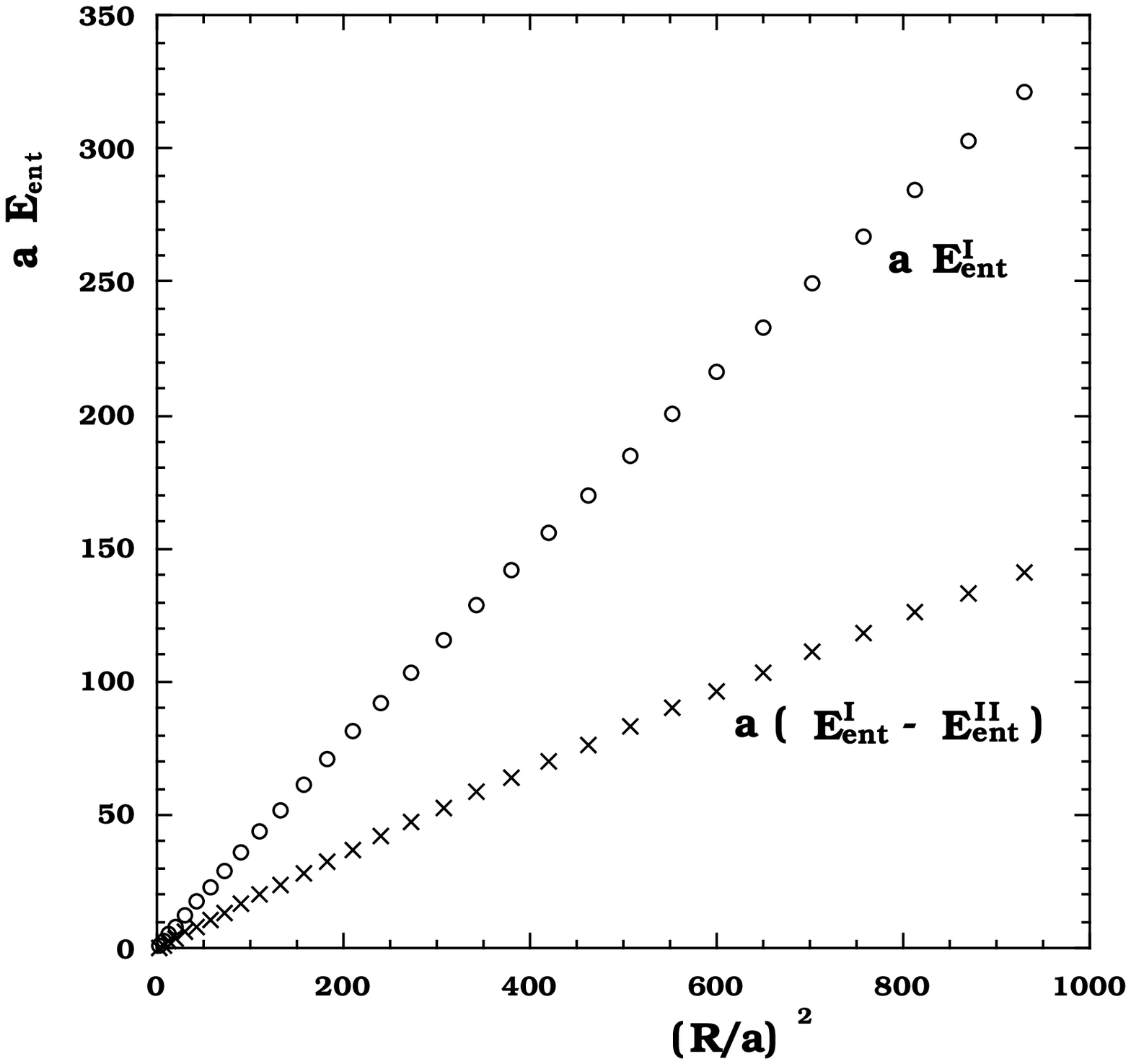}\vspace{1cm}}
 \caption{
The numerical evaluations for  
$aE_{ent}^I$ and $a(E_{ent}^I-E_{ent}^{II})$ 
for the  case of $B={\rm S}^2$. They are shown  as 
functions of $(R/a)^2$, where $R\equiv (n_B+1/2)a$. 
We have taken $N=60$ for
$aE_{ent}^I$ and $N=200$ for $a(E_{ent}^I-E_{ent}^{II})$. 
}
\label{fig:E-ent}
\end{figure}

Unfortunately, it is difficult to calculate $E_{ent}$ analytically. So 
we evaluated it by numerical calculation. First we evaluated
$E_{ent}^{(l)}$ for fixed values of $N$ and $n_B$ ($N>30$, 
$1\leq n_B\leq 30$). Next the summation with respect to $l$ was
performed up to $l=l_{max}$, which is determined so that 
\[
 (2l_{max}+1)E_{ent}^{(l_{max})}/\sum_{l=0}^{l_{max}}
	(2l+1)E_{ent}^{(l)}\leq	10^{-3}\ \ \ .
\]
Then we repeated the above procedure for all values of $n_B$ in the
range ($1\leq n_B\leq 30$). Finally we confirmed that $N=60$ for
$E_{ent}^I$ and $N=200$ for $E_{ent}^{I}-E_{ent}^{II}$ are so large
that the boundary condition $\phi_{lm,N+1}\equiv 0$ does not affect
the results. The results are shown in  {\it Figure} \ref{fig:E-ent},
where $aE_{ent}^I$ and $a(E_{ent}^{I}-E_{ent}^{II})$ are written as 
functions of $(R/a)^2=(n+1/2)^2$. 

From  this figure we see that $aE_{ent}$ is almost
proportional to $(R/a)^2$. Hence both $E_{ent}^I$ and $E_{ent}^{II}$
are proportional to $R^2/a^3$:
%
\begin{eqnarray}
 E_{ent}^{I} 
& \sim & 0.35\frac{R^2}{a^3}\ \ \ ,	\\
 E_{ent}^{II} 
& \sim & 0.20\frac{R^2}{a^3}\ \ \ .
\end{eqnarray}
These results again confirm  the relation Eq.(\ref{eqn:Eent-result}).


\section{Comparison: the thermodynamics of the entanglement 
and the black-hole thermodynamics} 
\label{sec:discussion}

\subsection{The thermodynamics of the entanglement in flat spacetime}
\label{subsection:thermo-ent}
We have introduced two possible definitions of entanglement 
energy, $E_{ent}^I$ and $E_{ent}^{II}$, in the previous 
section. By combining each of them  with the entanglement entropy 
$S_{ent}$, we obtain  two 
kinds of the thermodynamics of entanglement.

For this purpose, we consider an infinitesimal process in which 
the way of the division of the Hilbert space $\cal H$ 
into ${\cal H}_1$ and ${\cal H}_2$ is 
changed smoothly, with the `initial' state $\rho_0$ 
being fixed. 
(See  \ref{subsection:Senta}.)
 Let 
$dS_{ent}$ and $dE_{ent}$ be  
the  resultant infinitesimal changes in  the entanglement
entropy and in  the  entanglement energy, 
respectively. We  are dealing  with 
a 1-parameter family of the infinitesimal changes 
 for  the entanglement. The parameter 
 is chosen to be the area $A$ of $B$ for both of the 
 models in \ref{subsection:half} and in \ref{subsection:sphere}. 
  Thus, the construction of 
 the thermodynamics means to use (see Eq.(\ref{sec:Introduction})) 
%
\begin{equation}
  dE_{ent}=T_{ent} dS_{ent} 
\label{eqn:1st-law-ent}
\end{equation}
to determine $T_{ent}$, which is interpreted as the temperature of the
entanglement. Combining (\ref{eqn:Sent-result}) and
(\ref{eqn:Eent-result}) with Eq.(\ref{eqn:1st-law-ent}), we thus
get\footnote{In this section, we recover $\hbar$ and $c$.} 
%
\begin{equation}
 k_B T_{ent} = \hbar c \frac{{\cal N}_E/{\cal N}_S}{a} \ \ \ .
 \label{eqn:Tent}
\end{equation}
Note that the temperatures $T_{ent}$ obtained from the two definition
of the entanglement energy (Eq.(\ref{eqn:Eent1}) and
Eq.(\ref{eqn:Eent2})) coincides up to numerical factors of order
unity, because of the universal behavior of
Eq.(\ref{eqn:Eent-result}). 

Let us interpret the  thermodynamics of the  entanglement
given  by 
(\ref{eqn:Sent-result}), (\ref{eqn:Eent-result}) and
(\ref{eqn:Tent}). 
It is helpful  to introduce 
the quantities
%
\begin{eqnarray}
 n_{ent} & \equiv & \frac{A}{a^2} \ \ \ , \nonumber  \\
 e_{ent} & \equiv & \frac{\hbar c}{a} \ \ \ .
\label{eqn:unitenergy-ent}
\end{eqnarray}
Here $n_{ent}$ is regarded as an effective 
number of degrees of freedom of matter on the boundary $B$, and
$e_{ent}$ is a typical energy scale of each degree of freedom on $B$. 

From Eqs. (\ref{eqn:Sent-result}), (\ref{eqn:Eent-result}) and
(\ref{eqn:Tent}), we find that 
%
\begin{eqnarray}
 S_{ent}     & \sim & k_Bn_{ent} \ \ \ , \nonumber   \\
 E_{ent}     & \sim & e_{ent}n_{ent} \ \ \ , \nonumber   \\
 k_B T_{ent} & \sim & e_{ent} \ \ \ .
 \label{eqn:interpret-ent}
\end{eqnarray}
Therefore our results can be interpreted as follows\footnote{
It is safer, however, to regard such an interpretation just as a
convenient way of representing our results. This note in particular
applies to the case of the black-hole thermodynamics (see
Eq.(\ref{eqn:interpret-BH})).} 
: The entanglement entropy is a measure for the number of 
microscopic degrees of  freedom on the boundary $B$; 
the entanglement energy is a measure for  the 
total energy carried by all of the degrees of freedom on $B$; 
the temperature of the 
entanglement is measure for the energy carried by each 
degree of freedom on $B$.

\subsection{Discrepancy between the thermodynamics of the entanglement 
in flat spacetime and the black-hole thermodynamics} 
\label{subsection:discrepancy}
Now we compare these results with the case of black holes.
For that purpose we express the black-hole thermodynamics in the same
form as in the previous subsection.
   
Let us introduce the  quantities  
%
\begin{eqnarray}
 n_{BH} & \equiv & \frac{A}{l_p^2}\ \ \ , \nonumber	\\
 e_{BH} & \equiv & \frac{\hbar c}{l_{pl}}\ \ \ , 
\label{eqn:unitenergy-BH}
\end{eqnarray}
We can interpret that $n_{BH}$ corresponds to the effective number of
degrees of freedom on the event horizon  and $e_{BH}$ is a typical
energy scale  for each degree of freedom of matter on the horizon. 

The black-hole thermodynamics can be recast in terms of 
these quantities as 
%
\begin{eqnarray}
 S_{BH} & \sim & k_B n_{BH}\ \ \ , \nonumber  \\
 E_{BH} & \sim & \gamma_{BH}e_{BH} n_{BH}\ \ \ , \nonumber  \\
 k_B T_{BH} & \sim & \gamma_{BH}e_{BH}\ \ \ ,
\label{eqn:interpret-BH}
\end{eqnarray}
where $\gamma_{BH}\equiv l_{pl}/R$. The factor $\gamma_{BH}$ can be
understood as a magnification of energy due to an addition of
gravitational energy or a red-shift factor of temperature since
$\sqrt{-g_{tt}}\sim l_{pl}/R$ at $r\sim R+l_{pl}^2/R$, which
corresponds to a stationary observer at the proper distance $l_{pl}$
away from the horizon. Here $R$ is the area radius of the
horizon. Thus the following interpretation is possible\footnote{ 
See the footnote after Eq.(\ref{eqn:interpret-ent}).
}
: The black-hole entropy is a measure for the number of the
microscopic degrees of freedom  on the event horizon; 
the black-hole energy is a measure at infinity for the total energy
carried by all of the degrees of freedom on the event horizon; 
the black-hole temperature is a measure at infinity for the energy
carried by each degree of freedom.

Now we compare the two types of thermodynamics characterized by
Eq.(\ref{eqn:interpret-ent}) and Eq.(\ref{eqn:interpret-BH}),
respectively. Both of them allow the interpretation that they
describe the behavior of the effective microscopic degrees of freedom
on the boundary $B$, or on the horizon. Because of the factor
$\gamma_{BH}$, however, they are hardly understood in a unified
picture. This strongly suggests that an inclusion of gravitational
effects is necessary for agreement between them. In the remaining of
this subsection we shall see that the discrepancy cannot be avoided by
any means unless gravitational effects are taken into account for the
thermodynamics of the entanglement. After that, a restoration of the
agreement by gravity is discussed in the next subsection. 

The discrepancy is highlighted in the context of the third law of
thermodynamics. Both types of thermodynamics fail to follow the third
law (when $A$ is chosen as a control-parameter), but in quite
different manners. 
 
In Eq.(\ref{eqn:Tent}), 
we see that $T_{ent}$ remains constant if 
$\frac{{\cal N}_E}{{\cal N}_S}$ is assumed to be constant.\footnote{
Here we are regarding the cut-off scale $a$ as the fundamental
constant of the theory, not to be varied. However, see the discussions
below.}
On the other hand, Eq.(\ref{eqn:Sent-result}) shows that 
$S_{ent}$ tends to zero as $A \to 0$. Therefore we obtain the
following $A$-dependence:
%
\begin{eqnarray}
S_{ent} & \propto & A \ \ \ , \nonumber \\
E_{ent} & \propto & A \ \ \ , \nonumber \\
k_B  T_{ent}   & \propto & A^0 \ \ \ .
\label{eqn:ent-A}
\end{eqnarray}     
The system  behaves  as though it is kept in touch  with 
a thermal bath with temperature $T_{ent}$.

In contrast, for the black-hole thermodynamics, Eq.(\ref{eqn:T-BH})
and Eq.(\ref{eqn:S-BH}) along with Eq.(\ref{eqn:constants}) give the
behavior (note that $A \propto M_{BH}^2$) 
%
\begin{eqnarray}
S_{BH} & \propto & A \ \ \ , \nonumber \\
E_{BH} & \propto & \sqrt{A} \ \ \ , \nonumber \\
k_B  T_{BH}   & \propto & 1/ \sqrt{A} \ \ \ .
\label{eqn:BH-A}
\end{eqnarray}     
    Thus we see that $S_{BH} \to \infty$ as $T_{BH} \to 0$. 

The discrepancy between Eq.(\ref{eqn:ent-A}) and 
Eq.(\ref{eqn:BH-A}) is 
quite impressive. On one hand, 
a well-known behavior (\ref{eqn:BH-A}) comes from 
the fundamental properties of the black-hole physics. On the 
other hand,  
the behavior  characterized by Eq.(\ref{eqn:ent-A}) is also 
an  universal one  in any model of the entanglement: 
The zero-point energy of the system has been subtracted as  
Eq.(\ref{eqn:Eent1}) or Eq.(\ref{eqn:Eent2}), thus 
only the degrees of freedom on the boundary $B$ contributes 
to $E_{ent}$, yielding the behavior $E_{ent} \propto A$. 
The two definitions of $E_{ent}$ proposed here 
(Eq.(\ref{eqn:Eent1}) and Eq.(\ref{eqn:Eent2})) 
look quite reasonable though other definitions 
may  be possible. The result $E_{ent} \propto A$  also looks 
natural, being compatible with the concept of `entanglement'.
At the same time,  $S_{ent}$ also  behaves  universally as 
$S_{ent} \propto A$, which has been  the original motivation for
investigating the relation between  $S_{BH}$ and
$S_{ent}$~\cite{BLLS,Srednicki,CW,FN,Sent}.

It is also interesting to investigate the cut-off dependence of both
types of thermodynamics. From the viewpoint of the theory of
the renormalization group~\cite{MA}, this dependence also deserves to
be investigated. 
    
First, for the case of the thermodynamics of entanglement, 
$a$ should be  varied with  $A$ being  fixed. Hence, from 
Eqs.(\ref{eqn:Sent-result}), (\ref{eqn:Eent-result}) and 
(\ref{eqn:Tent}),  we see  that
%
\begin{eqnarray}
S_{ent} & \propto & a^{-2}\ \ \ , \nonumber \\
E_{ent} & \propto & a^{-3}\ \ \ , \nonumber \\
k_B  T_{ent}   & \propto & a^{-1}\ \ \ .
\label{eqn:ent-a}
\end{eqnarray}     
When we regard $a$ instead of $A$ as the external 
control-parameter of the system, thus, the third law of 
thermodynamics follows: $S_{ent} \to 0$ as $T_{ent} \to 0$. 

For the case of the black-hole thermodynamics,  on the other hand, 
$l_p$ should be varied with 
$A=\frac{16\pi G^2}{c^4} M_{BH}^2=16\pi {l_p}^4 
\left( \frac{\hbar}{M_{BH}c} \right)^{-2}$ being fixed.\footnote{
There is no direct connection between $a$ and $l_p$. However 
they have a common property that both of them 
introduces  a cut-off scale into  a quantum matter field.} 
Then we see from Eqs.(\ref{eqn:S-BH}) (with (\ref{eqn:constants})), 
(\ref{eqn:E-BH}) and (\ref{eqn:T-BH}) that 
%
\begin{eqnarray}
S_{BH} & \propto & {l_p}^{-2}\ \ \ , \nonumber \\
E_{BH} & \propto & {l_p}^{-2}\ \ \ , \nonumber \\
k_B T_{BH}    & \propto & {l_p}^{0}\ \ \ .
\label{eqn:BH-l}
\end{eqnarray}     
Thus the third law of  thermodynamics does not hold
in this case, too. It is interesting to note that 
this behavior looks  similar to that in the case 
of the entanglement with $A$ being varied 
(Eq.(\ref{eqn:ent-A})).

Finally, for completeness let us look at the behavior 
of thermodynamics of entanglement and the black-hole thermodynamics
when $M_{BH}$ and $E_{ent}$ are fixed, respectively. It becomes 
%
\begin{eqnarray}
S_{BH} & \propto & {l_p}^{2}\ \ \ , \nonumber \\
E_{BH} & \propto & {l_p}^{0}\ \ \ , \nonumber \\
k_B T_{BH}    & \propto & {l_p}^{-2}\ \ \ .
\label{eqn:BH-l'}
\end{eqnarray}     
and 
%
\begin{eqnarray}
S_{ent} & \propto & a\ \ \ , \nonumber \\
E_{ent} & \propto & a^{0}\ \ \ , \nonumber \\
k_B  T_{ent}   & \propto & a^{-1}\ \ \ .
\end{eqnarray}     
The third law fails to hold in this case.

The results Eq.(\ref{eqn:ent-A}) in comparison with
Eq.(\ref{eqn:BH-A}), and Eq.(\ref{eqn:ent-a}) in comparison with
Eq.(\ref{eqn:BH-l}) are summarized in {\it Table}
\ref{table:summary}. Anyway, the thermodynamics of the entanglement in 
flat spacetime shows significant differences from the black-hole
thermodynamics.

\subsection{Restoration of the agreement by gravity}

In this subsection let us discuss a possible restoration of the
agreement between the thermodynamics of entanglement and the
black-hole thermodynamics by considering gravitational effects. 

Although the behavior (\ref{eqn:interpret-ent}) of the thermodynamics
of the entanglement was derived by considering models in flat
spacetime, it seems very reasonable that we regard the quantities
$S_{ent}$, $E_{ent}$ and $T_{ent}$ as those in a black-hole background 
measured by a stationary observer located at the proper distance $a$
away from the horizon \footnote{
The authors thank T. Jacobson for helpful comments on this point.
}.
Since $S_{BH}$, $E_{BH}$ and $T_{BH}$ in (\ref{eqn:interpret-BH}) are
quantities measured at infinity, it is behavior of $S_{ent}$,
$E_{ent}$ and $T_{ent}$ at infinity that we have to compare with
(\ref{eqn:interpret-BH}). $S_{ent}$ at infinity probably has the same
behavior as that measured by the observer near the horizon since a
number of degrees of freedom seems independent of an observer's
view-point. That is consistent with the fact that the entanglement
entropy on Schwarzschild background has the same behavior 
$S_{ent}\sim k_Bn_{ent}$ \cite{FN}. On the other hand it seems
natural to add the gravitational energy to the entanglement energy by
replacing $E_{ent}$ with $\sqrt{-g_{tt}}E_{ent}$. Then the
entanglement temperature is determined by use of the first law
(\ref{eqn:1st-law-ent}). Thus the inclusion of gravity may alter the 
behavior (\ref{eqn:interpret-ent}) to 
%
\begin{eqnarray}
 S_{ent}     & \sim & k_Bn_{ent} \ \ \ , \nonumber   \\
 E_{ent}     & \sim & \gamma_{ent}e_{ent}n_{ent} \ \ \ , \nonumber   \\
 k_B T_{ent} & \sim & \gamma_{ent}e_{ent} \ \ \ ,
	\label{re-interpret-ent}
\end{eqnarray}
where $\gamma_{ent}\equiv a/R$. The factor $\gamma_{ent}$ represents
the gravitational magnification of the entanglement energy due to the 
addition of gravitational energy since on the corresponding
Schwarzschild background $\sqrt{-g_{tt}}\sim a/R$ at $r\sim R+a^2/R$,
which corresponds to a stationary observer at the proper distance $a$
away from the horizon (see the argument below
(\ref{eqn:interpret-BH}) and Ref.\cite{FN,cutoff}). Here $R$ is the
area radius of the horizon. (\ref{re-interpret-ent}) shows a complete
agreement with (\ref{eqn:interpret-BH}). Note that the last equality 
in  (\ref{re-interpret-ent}) is consistent with an interpretation that
the entanglement temperature is red-shifted by the factor
$\gamma_{ent}$. Thus the inclusion of gravitational effects restores 
the agreement between the thermodynamics of the entanglement and the
black-hole thermodynamics at least qualitatively.


\section{Summary and discussions}	
\label{sec:summary}

In this paper we have tried to judge whether 
the black-hole entropy can be understood  
as the entanglement entropy associated with the division of spacetime
by the event horizon. 
Our strategy has been  to look at  the whole thermodynamical 
structures inherent in a black-hole system  and 
models introduced in \cite{BLLS,Srednicki} to relate their
entanglement entropy to the black-hole entropy. 
Following this strategy we have undertaken 
the construction of  the thermodynamics of entanglement. 
For this purpose, after reviewing the basics of the entanglement
entropy $S_{ent}$~\cite{BLLS,Srednicki,CW,FN,Sent}, we have 
proposed two reasonable definitions 
of the entanglement energy $E_{ent}$. To obtain explicit values of
$E_{ent}$, we have prepared basic formulas for 
the entanglement energy. We have then estimated the 
entanglement energy  by choosing 
two tractable models of a scalar field in the Minkowski space.
The boundary $B$ has been chosen as, respectively, 
$B={\rm R}^2$ and $B={\rm S}^2$ in each model. We have thus 
found a common behavior independent of the 
definition of $E_{ent}$, 
$E_{ent} \propto A/{a^3}$, where $A$ is the area of 
the boundary and $a$ is a fundamental cut-off scale of the system. 
Getting $E_{ent}$
along with $S_{ent}$ in hand, we have 
constructed the thermodynamics of entanglement 
by postulating the first law of thermodynamics. 
In particular 
we have found that the temperature of the entanglement 
$T_{ent}$ is proportional to $1/a$. Finally we have 
compared the thermodynamics of entanglement with 
the black-hole thermodynamics from various angles.

Though both of them allow the interpretation that the degrees of
freedom around the boundary $B$ (or  the event horizon) carry the
thermodynamical properties\footnote{ 
Needless to say, this interpretation is nothing more than 
one concise way of grasping  the situations among 
many possibilities.
}, 
it seems quite difficult to find  further parallelism between them. The
difficulty becomes clear in the context of the third law of
thermodynamics. Namely both the response to the variation of  $A$
(with $a$ or $l_p$ being fixed) and the response to the variation of
$a$ or $l_p$ (with $A$ being fixed) are very different in these two
types of thermodynamics ({\it Table} \ref{table:summary}). As is
discussed in \S\ref{subsection:discrepancy}, this discrepancy is
expected to be a universal one, independent of the models to be
considered. As for a system of the entanglement, both  $S_{ent}$ and
$E_{ent}$ become proportional to the area of the boundary, $A$, by the
very nature of the entanglement. Then  we get the universal behaviors
of the thermodynamics of entanglement, Eq.(\ref{eqn:ent-A}). On the
other hand, the behaviors Eq.(\ref{eqn:BH-A}) of the black-hole
thermodynamics are   well-established results. Though $S_{BH}$ is also
proportional to the area of the event horizon, $A$, like $S_{ent}$,
$E_{BH}$ is proportional to $\sqrt{A}$ rather than  $A$. 

What is the reason of the discrepancy? One simple answer is that the
thermodynamics of the entanglement we obtained is in flat spacetime 
and does not include any effects of gravity. We have discussed a
possible restoration of the agreement between them by introducing 
gravitational effects. It is due to a magnification of the
entanglement energy by an addition of gravitational energy and looks
very reasonable for us. Thus we can expect that the thermodynamics of
the entanglement behaves like the black-hole thermodynamics if
gravitational effects are taken into account properly. Any way, the
entanglement entropy passes a non-trivial check to be the black-hole
entropy. Finally we mention that our expectation is based on a
qualitative argument and that more quantitative and detailed analysis
along this line is needed.  

It will be also valuable to analyze the cut-off dependence of the
thermodynamics of   entanglement more systematically from the
viewpoint of the renormalization group.

\vskip 1cm

\centerline{\bf Acknowledgments}
M.S. was supported by the Yukawa Memorial Foundation, the Japan
Association for Mathematical Sciences and the Japan Society for
Promotion of Science. H.K. was supported by the Grant-in-Aid for
Scientific Research (C) from the Ministry of Education, Science,
Sports and Culture of Japan (40161947).

\appendix
\section{Symmetric property of the entanglement entropy for a pure state}
In this appendix we first give an abstract expression for the reduced
density operators $\rho_{\cal V}$ and $\rho_{\cal W}$ corresponding to 
a pure state $u$ in ${\cal U}={\cal V}\bar{\otimes}{\cal W}$, which do 
not use the subtrace. Then with the help of them we prove that
$S_{ent}$ obtained from $\rho_{\cal V}$ and $\rho_{\cal W}$ coincide
with each other. We follow the notations in \S\ref{subsection:Senta}. 

%
%
\begin{prop}	\label{prop:A_u}
For an arbitrary element $u$ of ${\cal U}={\cal V}\bar{\otimes}{\cal
W}$, there are antilinear bounded operators $A_u$ 
$\in\bar{{\cal B}}({\cal V},{\cal W})$ and $A^*_u$ 
$\in\bar{{\cal B}}({\cal W},{\cal V})$ such that 
%
\begin{equation}
 (A_ux,y) = (A^*_uy,x) = (u,x\otimes y)	\label{eqn:A_u-A^*_u}
\end{equation}
for ${}^{\forall}x\in{\cal V}$ and ${}^{\forall}y\in{\cal W}$.
\end{prop}

%
%
{\it Proof.}
Fix an arbitrary element $x$ of ${\cal V}$. Then $(u,x\otimes y)$
gives a linear bounded functional of $y (\in{\cal W})$ since 
%
\[
 \left| (u,x\otimes y)\right| \leq ||u|| ||x|| ||y||.
\]
Hence by Riesz's theorem there is a unique element $z_{u,x}$ of 
${\cal W}$ such that 
%
\begin{equation}
 (z_{u,x},y) = (u,x\otimes y)
\end{equation}
for ${}^{\forall}y\in{\cal W}$. 
Let us define $A_u$ by $A_u:x\to z_{u,x}$. It is evident that $A_u$
is an antilinear bounded operator from ${\cal V}$ to ${\cal W}$ since
%
\[
 ||A_{u}x|| = ||z_{u,x}|| = ||u|| ||x||.
\]

Exchanging the roles played by ${\cal V}$ and ${\cal W}$ in the above
argument, it is shown that there is an antilinear bounded operator 
$A^*_u$ from ${\cal W}$ to ${\cal V}$ such that 
$(A^*_uy,x) = (u,x\otimes y)$.
\hfill $\Box$

Note that $A_u$ and $A^*_u$ defined above are written as 
%
\begin{eqnarray}
 A_ux & = & \sum_jf_j(x\otimes f_j,u)\ \ \ ,	\nonumber\\
 A^*_uy & = & \sum_ie_i(e_i\otimes y,u)\ \ \ .
\end{eqnarray}
Using this expression, it is easily shown that 
%
\begin{eqnarray}
 A^*_uA_ux & = & \sum_{ij}e_i\left( e_i\otimes f_j,
			(u,x\otimes f_j)u\right)\ \ \ ,	\nonumber\\
 A_uA^*_uy & = & \sum_{ij}f_j\left( e_i\otimes f_j,
			(u,e_i\otimes y)u\right)\ \ \ .
\end{eqnarray}
These coincide with $\rho_{\cal V}$ and $\rho_{\cal W}$, respectively,
if $u$ has unit norm (see Eq. (\ref{eqn:rho}) and (\ref{eqn:reduced})). 
Therefore the following proposition says that $\rho_{\cal V}$ and
$\rho_{\cal W}$ have the same spectrum and the same multiplicity and
that entropy of them are identical.

%
%
\begin{prop}
$\rho_{u,{\cal V}}$ $(\in{\cal B}({\cal V}))$ and 
$\rho_{u,{\cal W}}$ $(\in{\cal B}({\cal W}))$ defined by
%
\begin{eqnarray}
 \rho_{u,{\cal V}} & = & A^*_uA_u\ \ \ ,\nonumber\\
 \rho_{u,{\cal W}} & = & A_uA^*_u\ \ \ 
\end{eqnarray}
are non-negative, trace-class self-adjoint operators, where $A_u$ and
$A^*_u$ are defined in Proposition \ref{prop:A_u} for an arbitrary
element $u$ of ${\cal U}$. The spectrum and the multiplicity of
$\rho_{u,{\cal V}}$ and $\rho_{u,{\cal W}}$ are identical for all
non-zero eigenvalues. 
\end{prop}

%
%
{\it Proof.}
In general
%
\[
 (x',\rho_{u,{\cal V}}x) = (A_ux,A_ux')
\]
for ${}^{\forall}x,x'$ ($\in{\cal V}$) by definition. Therefore 
%
\begin{equation}
 (x,\rho_{u,{\cal V}}x) = ||A_ux||^2 \geq 0
\end{equation}
and 
%
\begin{eqnarray}
 {\rm Tr}_{\cal V}(\rho_{u,{\cal V}}) 
	& = & \sum_i(A_ue_i,A_ue_i)	\nonumber\\
	& = & \sum_{i,j}(A_ue_i,f_j)(f_j,A_ue_i)	\nonumber\\
	& = & \sum_{i,j}|(u,e_i\otimes f_j)|^2	\nonumber\\
	& = & ||u||^2,
\end{eqnarray}
i.e. $\rho_{u,{\cal V}}$ is non-negative and trace-class. In general a
non-negative operator is self-adjoint and a trace-class operator is
compact. Thus the eigenvalue expansion theorem for a self-adjoint
compact operator says that all eigenvalues of $\rho_{u,{\cal V}}$ are
discrete except zero and have finite multiplicity. For a later
convenience let us denote the non-zero eigenvalues and the
corresponding eigenspaces as $\lambda_i$ and ${\cal V}_i$
($i=1,2,\cdots $). 

Similarly, it is shown that $\rho_{u,{\cal W}}$ is non-negative and
trace-class and that all eigenvalues of it are discrete except zero
and have finite multiplicity.

Now $\ker\rho_{u,{\cal W}}=\ker A^*_u$ since 
$\rho_{u,{\cal W}}y=A_uA^*_uy$ and 
$(y,\rho_{u,{\cal W}}y)=||A^*_uy||^2$ for an arbitrary element $y$
of ${\cal W}$ by definitions. Moreover, from (\ref{eqn:A_u-A^*_u}) it
is evident that  
%
\[
 y\perp\mbox{Ran}A_u \Leftrightarrow y\in\ker A^*_u.
\]
With the help of these two facts ${\cal W}$ is decomposed as
%
\begin{equation}
 {\cal W} = \ker \rho_{u,{\cal W}}\oplus\overline{\mbox{Ran}A_u}.
	\label{eqn:W=ker+ran}
\end{equation}
where the overline means to take a closure.

Moreover, it is easily shown by definitions that 
%
\begin{eqnarray}
 \rho_{u,{\cal W}}A_ux
	& = & \lambda_iA_ux\ \ \ ,\nonumber\\
 \left( A_ux,A_ux'\right) & = & \lambda_i(x',x)
\end{eqnarray}
for ${}^{\forall}x$ ($\in{\cal V}_i$) and 
${}^{\forall}x'$ ($\in{\cal V}_{i'}$). Hence $A_u$ maps the eigenspace
${\cal V}_i$ to a eigenspace of $\rho_{u,{\cal V}}$ with the 
same eigenvalue, preserving its dimension. Taking account of
Eq.(\ref{eqn:W=ker+ran}), this implies that the spectrum and the 
multiplicity of $\rho_{u,{\cal V}}$ and $\rho_{u,{\cal W}}$ are
identical for all non-zero eigenvalues.
\hfill $\Box$


\newpage

\newpage

%
\begin{table}
\caption{Comparison of two kinds of thermodynamics}
\label{table:summary}
 \begin{center}
  \begin{tabular}{rcc} \hline
		& {\it Entanglement in flat spacetime}	
		& {\it Black-Hole} \\ \hline
   {\it Varied}	& $A$			& $A$ \\
   {\it Fixed}	& $a$			& $l_{pl}$ \\
	$S$	& $\propto A$		& $\propto A$ \\
	$E$	& $\propto A$		& $\propto A^{1/2}$ \\
	$T$	& $\propto A^0$		& $\propto A^{-1/2}$ \\ \hline
   {\it Varied}	& $a$			& $l_{pl}$ \\
   {\it Fixed}	& $A$			& $A$ \\
	$S$	& $\propto a^{-2}$	& $\propto l_{pl}^{-2}$	\\
	$E$	& $\propto a^{-3}$	& $\propto l_{pl}^{-2}$	\\
	$T$	& $\propto a^{-1}$	& $\propto l_{pl}^{0}$ \\ 
								\hline 
   {\it Varied}	& $a$			& $l_{pl}$ \\
   {\it Fixed}	& $E_{ent}$		& $M$ \\
	$S$	& $\propto a$		& $\propto l_p^2$ \\
	$E$	& $\propto a^0$		& $\propto l_p^0$ \\
	$T$	& $\propto a^{-1}$	& $\propto l_p^{-2}$ \\ \hline
    \end{tabular}
 \end{center}
\end{table}

\end{document}